\documentclass[aps,prd,preprintnumbers,nofootinbib,superscriptaddress]{revtex4-2}

\usepackage{amsmath,amssymb}
\usepackage{graphicx}
\usepackage{bm}
\usepackage{hyperref}

\begin{document}

\title{
Evidence for Log-Periodic Modulation in the Redshift Distribution
of High-Redshift Compact Sources
}

\author{Takeshi Fukuyama}
\affiliation{
Research Center for Nuclear Physics (RCNP),
Osaka University, Ibaraki, Osaka 567-0047, Japan
}

\begin{abstract}
We investigate whether the redshift distribution of high-redshift
compact sources identified in early James Webb Space Telescope
(JWST) observations contains a statistically coherent periodic
component in the logarithmic redshift variable
$x=\ln(1+z)$. Using a publicly available catalog of 341 sources,
we perform an unbinned frequency analysis in which the smooth
redshift dependence is removed using a baseline probability
distribution. The resulting baseline-corrected power spectrum
exhibits a prominent feature at
$\omega_{\rm peak}=18.33$, corresponding to a logarithmic spacing
$\Delta x\simeq0.34$, or a multiplicative spacing of approximately
1.41 in $1+z$. Monte Carlo realizations of the smooth null model
give a local tail probability
$p_{\rm local}\simeq2.8\times10^{-3}$. Accounting for the
frequency scan over the theoretically motivated interval
$16\leq\omega\leq21$ gives a global false-alarm probability
$p_{\rm global}\simeq7.6\times10^{-3}$.
Cross-validation, repeated train/test holdout analyses, and
field-dependent tests are used to examine the robustness of the
feature against statistical fluctuations and sample heterogeneity.
The statistical analysis is independent of the physical nature of
the individual compact sources. If confirmed with larger and more
homogeneous samples, an approximately constant spacing in
$\ln(1+z)$ would indicate a preferred multiplicative scale in the
high-redshift source distribution. Recurrent cosmological
condensate dynamics provide one possible physical interpretation,
but the statistical result itself does not depend on this mechanism.
\end{abstract}

\maketitle

\section{Introduction}
\label{sec:intro}

The James Webb Space Telescope (JWST) has substantially extended
observations of galaxies and compact sources into the high-redshift
Universe.  Among the unexpected results is the identification of a
population of compact red sources, often referred to as Little Red
Dots (LRDs), whose abundance and spectral properties have attracted
considerable attention
\cite{Matthee2024,Greene2024,Kokorev2024}.
Photometric surveys have revealed a substantial population of such
compact sources over approximately $4<z<9$, while spectroscopic
observations have identified broad Balmer emission lines in many
objects, suggesting an important contribution from accreting black
holes.
The physical nature of LRDs, however, remains under active
investigation \cite{InayoshiHo2026,Rusakov2026}.
Their observed spectral energy distributions may contain
contributions from active galactic nuclei, compact stellar
populations, or both.
  Moreover, the inferred abundance of LRDs
depends to some extent on the adopted photometric and morphological
selection criteria.  It is therefore useful to distinguish questions
concerning the astrophysical identity of individual objects from
questions concerning statistical structure in their observed
redshift distribution.

The purpose of the present work is specifically to address the latter
question.  We ask whether the {\it redshift distribution} of the
currently available high-redshift compact-source sample contains a
coherent structure that cannot be readily described as a smooth
function of redshift.  This question can be posed independently of
whether every object in the catalog ultimately receives the same
astrophysical classification.

A particularly useful variable for this purpose is
\begin{equation}
 x \equiv \ln(1+z).
 \label{eq:xdef}
\end{equation}
A periodic modulation in $x$ has a simple geometrical meaning.
If successive features occur at
\begin{equation}
 x_{n+1}-x_n=\Delta x,
 \label{eq:spacing}
\end{equation}
then
\begin{equation}
 \frac{1+z_{n+1}}{1+z_n}
       = \exp(\Delta x),
 \label{eq:scaling}
\end{equation}
so that the corresponding sequence is characterized by a preferred
multiplicative scale rather than by a preferred additive interval in
redshift.  Such behavior is conventionally described as log-periodicity
and is closely related to discrete scale invariance
\cite{Sornette1998,JogiSornette1998,Huang1997}.

Interestingly, a logarithmic recurrence in redshift was proposed
considerably earlier in the context of cosmological Bose--Einstein
condensation (BEC).  In the relativistic Gross--Pitaevskii framework,
repeated condensation and rapid collapse can generate localized
structures at successive cosmological epochs
\cite{FukuyamaMorikawa2006,FukuyamaMorikawaTatekawa2008}.
The possible cosmological role of such condensate dynamics was
subsequently investigated in several related settings
\cite{FukuyamaMorikawa2009,Fukuyama2026}.

In the present work, however, we deliberately separate this possible
physical interpretation from the observational test.  The
observational question can be stated in a model-independent form:
\begin{quote}
{\it Does the observed distribution of high-redshift compact sources
contain a statistically significant periodic component in
$x=\ln(1+z)$ after the smooth redshift dependence has been removed?}
\end{quote}

To answer this question we use an unbinned frequency statistic.
For a trial angular frequency $\omega$, we define two
baseline-corrected components,
\begin{align}
 U(\omega)
 &=
 \sum_{i=1}^{N}\cos(\omega x_i)
 -
 N\left\langle\cos(\omega x)\right\rangle_0 ,
 \label{eq:Uintro}
 \\
 V(\omega)
 &=
 \sum_{i=1}^{N}\sin(\omega x_i)
 -
 N\left\langle\sin(\omega x)\right\rangle_0 ,
 \label{eq:Vintro}
\end{align}
where $\langle\cdots\rangle_0$ denotes an expectation value with
respect to a smooth baseline distribution $p_0(x)$,
\begin{equation}
\left\langle\cos(\omega x)\right\rangle_0=\int dx p_0(x)\cos (\omega x).
\end{equation}
The corresponding phase-independent power is
\begin{equation}
 P(\omega)
 =
 \frac{U^2(\omega)+V^2(\omega)}{N}.
 \label{eq:Pintro}
\end{equation}
Thus $U$ and $V$ measure the cosine and sine components of a possible
periodic modulation, while $P$ measures its strength without requiring
the phase to be specified in advance.

Applied to the current sample, this procedure identifies a
prominent feature near
\begin{equation}
 \omega \simeq 18.3,
\end{equation}
corresponding to
\begin{equation}
 \Delta x = \frac{2\pi}{\omega} \simeq 0.34.
\end{equation}
The statistical significance of this feature is evaluated by
Monte Carlo simulations, as described in detail in Sec.~III.C.
We therefore assess the statistical evidence in successive stages.
Figure~1 establishes the significance of the observed feature against
the smooth null model, while Figs.~2 and 3 test whether the selected
frequency retains statistical strength in independent subsets and
under repeated train/test partitions of the data.

An important issue is whether an apparent periodicity could be
generated by observational selection, heterogeneous survey fields,
redshift uncertainties, or contamination of the high-redshift sample.
We treat these possibilities as systematic questions rather than
assuming that all cataloged objects form a physically homogeneous
population.  In particular, the interpretation of individual compact
red sources may evolve as additional photometric and spectroscopic
observations become available.  The statistical claim investigated
here concerns the redshift distribution of the selected sample and
must therefore be tested for robustness against changes in sample
definition.

Only after establishing the statistical result do we consider its
possible physical interpretation.  A periodic structure in
$\ln(1+z)$ can arise when an underlying dynamical process possesses
an approximately recurrent scale.  One possible realization is
provided by cosmological BEC dynamics with attractive
self-interaction, in which repeated collapse and re-formation episodes
can occur
\cite{FukuyamaMorikawa2006,FukuyamaMorikawaTatekawa2008,
Fukuyama2026}.
We emphasize, however, that this interpretation is not assumed in
constructing the statistic and that the observation of a
log-periodic component, by itself, does not uniquely identify its
physical origin.

The organization of this paper is as follows.
Section~\ref{sec:data} describes the high-redshift source sample and
the construction of the logarithmic redshift variable.
Section~\ref{sec:method} defines the baseline-corrected unbinned
frequency statistic and the null hypothesis.
Section~\ref{sec:results} presents the frequency scan and its
local and global statistical significance.
Section~\ref{sec:robustness} examines cross-validation, field dependence,
and possible observational systematics.
Section~\ref{sec:interpretation} discusses possible physical interpretations,
including cyclic condensate dynamics.
Our conclusions are summarized in Sec.~\ref{sec:conclusions}.


\section{Data and sample definition}
\label{sec:data}

\subsection{High-redshift compact-source sample}

The data used in the present analysis are taken from the
publicly available catalog of Little Red Dots compiled by
Kocevski et al.~\cite{Kocevski2025}. The catalog contains
341 sources selected from the CEERS, PRIMER, JADES, UNCOVER,
and NGDEEP JWST surveys.

We use the full version of Table~3 made publicly available
in the associated GitHub repository~\cite{KocevskiCatalog}.
For each source, we adopt the best-estimate redshift
$z_{\rm best}$ listed in the catalog. The public table
contains 341 entries, and no additional object selection is
applied in constructing the working sample used here. Thus,
\begin{equation}
 N=341.
\end{equation}
The present frequency analysis uses only the redshift
information from the catalog; the other photometric
quantities are not used.

Although the sources are commonly classified as Little Red
Dots, their detailed physical interpretation remains under
discussion. The present analysis therefore treats them as an
observationally selected sample of compact high-redshift
sources and does not require a unique physical origin for
every object.

\subsection{Logarithmic redshift variable}

For each source with redshift $z_i$, we introduce the logarithmic
variable
\begin{equation}
 x_i = \ln(1+z_i).
 \label{eq:xdata}
\end{equation}
This transformation is central to the present analysis.
A periodic structure in $x$ corresponds to a multiplicative, rather
than additive, sequence in $1+z$.

Indeed, if two successive features satisfy
\begin{equation}
 x_{n+1}-x_n=\Delta x,
\end{equation}
then
\begin{equation}
 \frac{1+z_{n+1}}{1+z_n}
 =
 \exp(\Delta x).
 \label{eq:multiplicative}
\end{equation}
Thus a constant interval in $x$ represents a preferred scaling ratio
in $1+z$.
This is the form expected for a log-periodic signal or, more
generally, for a system exhibiting approximate discrete scale
invariance \cite{Sornette1998}.

The observed redshifts are therefore mapped directly into the set
\begin{equation}
 \{z_i\}_{i=1}^{N}
 \quad\longrightarrow\quad
 \{x_i=\ln(1+z_i)\}_{i=1}^{N},
\end{equation}
and all frequency analyses described below are performed in the
$x$ variable.

\subsection{Heterogeneity and observational uncertainties}

The use of a combined JWST sample requires particular care because
the individual survey fields differ in depth, area, source-selection
criteria, and redshift determination.
In addition, some objects have photometric redshift estimates whereas
others may have spectroscopic constraints.

Such effects can alter the smooth redshift distribution and, in
principle, may also generate artificial structure if not properly
controlled \cite{Zhang2025,Rinaldi2026}. For this reason we do not compare the observed distribution with a
uniform distribution in either $z$ or $x$.
Instead, the slowly varying part of the observed distribution is
represented by a smooth baseline probability density $p_0(x)$,
introduced in Sec.~\ref{sec:method}.

The statistical question is consequently not whether the raw source
counts are uniform in $x$, but whether a coherent oscillatory
component remains after subtraction of the smooth baseline.

Possible dependence on survey field, sample definition, and redshift
uncertainties is treated separately as a robustness question in
Sec.~\ref{sec:robustness}.
This separation is important because the existence of a peak in a
frequency statistic alone does not establish that the feature is
cosmological.

\subsection{Scope of the present analysis}

The present work is designed as a test of structure in the observed
redshift distribution.
It does not require that every source currently classified as an LRD
retain that classification under future observations.

If a small fraction of individual objects are subsequently
reclassified, the relevant question is whether removal or replacement
of such objects significantly changes the frequency-space feature
identified below.
This motivates the subsample, cross-validation, and field-dependent
tests discussed in Sec.~\ref{sec:robustness}.

\section{Statistical method}
\label{sec:method}

\subsection{Smooth null distribution}

The observed distribution of
\begin{equation}
 x=\ln(1+z)
\end{equation}
is not expected to be uniform.
Cosmic evolution, survey sensitivity, and source-selection effects
can all produce a slowly varying dependence on $x$.
It is therefore necessary to distinguish such a smooth component
from the oscillatory component that we wish to test.

We describe the smooth null distribution by a normalized probability
density $p_0(x)$,
\begin{equation}
 \int_{x_{\rm min}}^{x_{\rm max}} p_0(x)\,dx=1.
 \label{eq:p0norm}
\end{equation}
For an arbitrary function $f(x)$, we denote its expectation value
under the null distribution by
\begin{equation}
 \langle f(x)\rangle_0
 =
 \int_{x_{\rm min}}^{x_{\rm max}}
 dx\,p_0(x)f(x).
 \label{eq:nullaverage}
\end{equation}
The role of $p_0(x)$ is only to describe the slowly varying part of
the observed distribution.  It is not intended to contain a
log-periodic component.
For the fiducial analysis, the observed range in $x$ is
divided into ten equal-width bins.  We fit a cubic polynomial
to the logarithm of the binned source density.  The fitted
function is then exponentiated and normalized over the
observed range to define the smooth baseline probability
density $p_0(x)$.  This construction ensures a positive,
smooth null distribution while retaining only the slowly
varying component of the observed redshift distribution.
The dependence of the results on the baseline
parametrization is examined separately in Sec.~V.A.
Under the null hypothesis, $N$ independent objects drawn from
$p_0(x)$ have the expected Fourier components
\begin{align}
 \left\langle
 \sum_{i=1}^{N}\cos(\omega x_i)
 \right\rangle_0
 &=
 N\langle\cos(\omega x)\rangle_0,
 \label{eq:cosmean}
 \\
 \left\langle
 \sum_{i=1}^{N}\sin(\omega x_i)
 \right\rangle_0
 &=
 N\langle\sin(\omega x)\rangle_0.
 \label{eq:sinmean}
\end{align}
The factor $N$ is essential because the expectation value of the
sum is the sum of the expectation values of the $N$ independent
objects.

\subsection{Baseline-corrected Fourier components}
A possible log-periodic modulation in the distribution of sources
may be represented schematically as
\begin{equation}
 p(x) \simeq p_0(x)
 \left[1+A\cos(\omega x+\phi)\right],
 \qquad x=\ln(1+z),
 \label{eq:logperiodic}
\end{equation}
where $p_0(x)$ denotes the smooth underlying distribution,
$A$ is the modulation amplitude, $\omega$ is the angular frequency
in logarithmic redshift, and $\phi$ is an arbitrary phase.
The corresponding period in $x$ is
\begin{equation}
 \Delta x=\frac{2\pi}{\omega}.
 \label{eq:period}
\end{equation}
Thus, a modulation with $\omega=18.3$ corresponds to
$\Delta x\simeq0.34$, or equivalently to a multiplicative spacing
$(1+z_{n+1})/(1+z_n)\simeq e^{0.34}\simeq1.41$.
Equation~(\ref{eq:logperiodic}) is intended only to illustrate the
form of a possible log-periodic component; the analysis below does
not assume a particular value of the phase $\phi$.
For each trial angular frequency $\omega$, we subtract the expected
smooth contribution and define
\begin{align}
 U(\omega)
 &=
 \sum_{i=1}^{N}\cos(\omega x_i)
 -
 N\langle\cos(\omega x)\rangle_0,
 \label{eq:U}
 \\
 V(\omega)
 &=
 \sum_{i=1}^{N}\sin(\omega x_i)
 -
 N\langle\sin(\omega x)\rangle_0.
 \label{eq:V}
\end{align}

Thus $U(\omega)$ and $V(\omega)$ are the cosine and sine quadratures
of the residual distribution after subtraction of the Fourier
components expected from the smooth baseline.

This subtraction is important because even a completely smooth but
nonuniform distribution can possess nonzero Fourier components.
Without the subtraction, such components could be mistaken for an
oscillatory signal.

We combine the two quadratures into the phase-independent statistic
\begin{equation}
 P(\omega)
 =
 \frac{U^2(\omega)+V^2(\omega)}{N}.
 \label{eq:power}
\end{equation}
The normalization by $N$ makes $P(\omega)$ an intensive measure of
the coherent Fourier power and facilitates comparison with null
catalogs having the same number of objects.

To see why the combination $U^2+V^2$ is useful, consider schematically
a weak modulation of the form
\begin{equation}
 p(x)
 \simeq
 p_0(x)
 \left[
 1+A\cos(\omega x+\phi)
 \right],
 \label{eq:modulation}
\end{equation}
where $A$ and $\phi$ denote its amplitude and phase.
Changing $\phi$ redistributes the signal between the cosine and sine
quadratures, whereas the combination $U^2+V^2$ is insensitive to this
choice of phase.
Consequently, $P(\omega)$ tests for a periodic component without
requiring its phase to be specified in advance.

\subsection{Null hypothesis and Monte Carlo calibration}

The null hypothesis $H_0$ is that the observed values $x_i$ are
independent realizations of the smooth distribution $p_0(x)$, with
no additional coherent log-periodic component.

To calibrate the statistic, Monte Carlo catalogs containing the same
number of objects as the observed sample are generated from
$p_0(x)$.  For each simulated catalog the same statistic
$P_{\rm MC}(\omega)$ is evaluated.

At a specified frequency $\omega_\ast$, the fixed-frequency
tail probability is defined by
\begin{equation}
 p_{\rm fixed}
 =
 {\rm Prob}_{H_0}
 \left[
 P_{\rm MC}(\omega_\ast)
 \geq
 P_{\rm obs}(\omega_\ast)
 \right].
 \label{eq:pfixed}
\end{equation}
In practice, for $N_{\rm MC}$ Monte Carlo realizations this quantity
is estimated as
\begin{equation}
 p_{\rm fixed}
 \simeq
 \frac{N_{\rm exceed}}{N_{\rm MC}},
 \label{eq:pMC}
\end{equation}
where $N_{\rm exceed}$ is the number of simulated catalogs for which
the power at $\omega_\ast$ equals or exceeds the observed value.

The cyclic-collapse phenomenology motivates a characteristic
spacing of order
\begin{equation}
 \Delta x = O(0.3-0.4),
 \label{eq:band}
\end{equation}
which corresponds to the frequency interval
\begin{equation}
 \omega \sim \frac{2\pi}{\Delta x}
 \simeq 16-21.
 \label{eq:motivatedband}
\end{equation}
Within this theoretically motivated frequency band, the observed
power spectrum exhibits its dominant enhancement near
\begin{equation}
 \omega_{\rm peak} \simeq 18.33.
 \label{eq:peak}
\end{equation}
The statistical significance of this enhancement is calibrated
using Monte Carlo catalogs generated from the same fiducial
smooth null distribution $p_0(x)$.  The local tail probability
is evaluated at the frequency of the observed feature, while
the global false-alarm probability associated with scanning
the motivated frequency interval is evaluated separately in
Sec.~III.D.

\subsection{Frequency scan and the trials factor}

The cyclic-collapse phenomenology motivates the restricted
frequency interval
\begin{equation}
 16 \leq \omega \leq 21 .
\end{equation}
Within this interval, the observed power spectrum has its
maximum at
\begin{equation}
 \omega_{\rm peak}=18.33 .
\end{equation}
The corresponding observed power is
\begin{equation}
 P_{\rm obs}(\omega_{\rm peak})=5.87 .
\end{equation}

We distinguish the local tail probability from the global
false-alarm probability associated with the frequency scan \cite{GrossVitells2010}.
The local probability is evaluated at the frequency of the
observed feature. To account for the look--elsewhere effect,
for each Monte Carlo catalog we additionally compute
\begin{equation}
 P_{\max}^{\rm MC}
 =
 \max_{16\leq\omega\leq21} P_{\rm MC}(\omega).
\end{equation}
The global probability is then defined as
\begin{equation}
 p_{\rm global}
 =
 {\rm Prob}_{H_0}
 \left[
 P_{\max}^{\rm MC}
 \geq P_{\max}^{\rm obs}
 \right].
\end{equation}
Using $4\times10^4$ Monte Carlo catalogs generated from the
fiducial smooth null distribution, we obtain
\begin{equation}
 p_{\rm local}\simeq2.8\times10^{-3},
 \qquad
 p_{\rm global}\simeq7.6\times10^{-3}.
\end{equation}
Thus the look--elsewhere correction increases the false-alarm
probability, as expected, but the global probability remains
below one percent within the theoretically motivated
frequency interval.

\section{Evidence for a log-periodic component}
\label{sec:results}

\subsection{Frequency scan}

We now apply the baseline-corrected statistic defined in Sec.~III
to the observed sample of $N=341$ high-redshift compact sources.
The power $P(\omega)$ is evaluated as a function of the trial
frequency $\omega$.

As discussed in Sec.~III, the cyclic-collapse phenomenology motivates
a characteristic logarithmic spacing
\begin{equation}
 \Delta x = O(0.3-0.4),
\end{equation}
corresponding approximately to the frequency interval
\begin{equation}
 \omega \simeq 16-21.
\end{equation}
Within this motivated interval, the observed power spectrum
exhibits its dominant enhancement at
$\omega_{\rm peak}=18.33$, as given in Eq.~(\ref{eq:peak}).
%

The corresponding logarithmic spacing is
\begin{equation}
 \Delta x
 =
 \Delta\ln(1+z)
 =
 \frac{2\pi}{\omega_{\rm peak}}
 \simeq 0.34.
 \label{eq:deltaxobs}
\end{equation}
Equivalently, this spacing corresponds to the multiplicative ratio
\begin{equation}
 \frac{1+z_{n+1}}{1+z_n}
 =
 e^{\Delta x}
 \simeq 1.41.
 \label{eq:scaleratio}
\end{equation}
Thus, if the feature represents a genuine log-periodic component,
successive structures would be separated not by a constant additive
interval in redshift, but approximately by a constant scaling factor
in $1+z$.

The identification of $\omega_{\rm peak}\simeq18.33$ is based on the
frequency scan itself.  It should therefore be distinguished from
a frequency specified independently of the data, as emphasized in
Sec.~III.D. 
The statistical assessment proceeds in three stages.
Figure~1 compares the observed power with Monte Carlo
realizations of the smooth null model and provides the
basic significance calibration.  Figure~2 then examines
whether the selected frequency remains predictive when
determined from a training subset and evaluated on an
independent test subset.  Finally, Figure~3 repeats this
procedure over many random partitions of the data to test
the robustness of the result against the particular choice
of train/test split.
\begin{figure}[t]
\centering
\includegraphics[width=0.65\columnwidth]{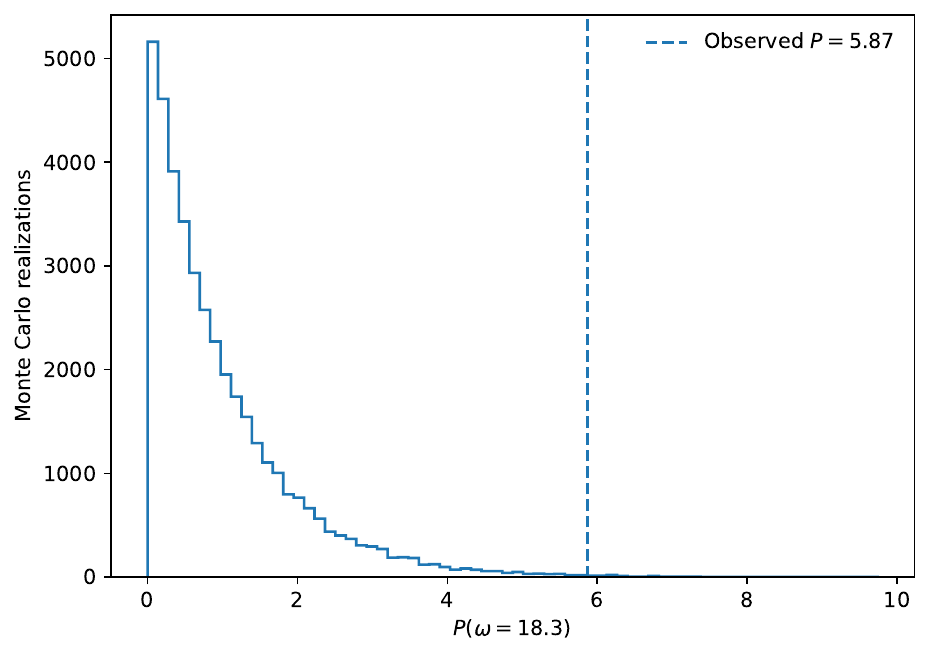}
\caption{Monte Carlo null distribution of the baseline-corrected
power $P(\omega)$ evaluated at $\omega=18.3$. The histogram shows
the distribution obtained from $4\times10^4$ realizations drawn
from the fiducial smooth baseline model $p_0(x)$. The vertical
line indicates the observed value $P_{\rm obs}=5.87$ from the
JWST data. The corresponding local tail probability is
$p_{\rm local}=2.8\times10^{-3}$.}
\label{fig:MC}
\end{figure}

\subsection{Local and global significance of the dominant feature}

At the location of the dominant enhancement,
$\omega_{\rm peak}=18.33$, the observed power is
\begin{equation}
 P_{\rm obs}(\omega_{\rm peak})=5.87 .
\end{equation}
Using $4\times10^4$ Monte Carlo catalogs generated from the
fiducial smooth null distribution, the fixed-frequency tail
probability evaluated at $\omega=18.3$, effectively the location
of the observed feature, is
\begin{equation}
 p_{\rm local}\simeq2.8\times10^{-3}.
\end{equation}
Because the peak frequency is identified from the frequency
scan, the local probability is not by itself the appropriate
false-alarm probability for the search.

We therefore repeat the frequency scan over
$16\leq\omega\leq21$ for every Monte Carlo catalog and
compare the maximum simulated power with the maximum
observed power.  This gives
\begin{equation}
 p_{\rm global}\simeq7.6\times10^{-3}.
\end{equation}
The look--elsewhere correction therefore weakens the nominal
significance, as expected, but the probability of obtaining
an equally large or larger peak somewhere within the
motivated frequency interval remains below one percent under
the specified smooth null model.

The statistical relevance of the feature also depends on
whether its location and amplitude are stable under changes
of the data sample and analysis procedure.  These questions
are addressed by the robustness and cross-validation tests
in the following section.

\section{Robustness and possible systematics}
\label{sec:robustness}

The statistical significance obtained in Sec.~IV does not by itself
establish that the feature near $\omega\simeq18.3$ is robust.
In particular, an apparent enhancement could depend on the choice
of the smooth baseline, on a particular subset of objects, or on
one of the survey fields.  We therefore perform several tests in
which the analysis is repeated after modifying either the statistical
procedure or the data subset.

\subsection{Dependence on the smooth baseline}

The smooth baseline distribution $p_0(x)$ is introduced to
describe the slowly varying redshift dependence that is not
associated with the oscillatory component being tested.
Since the baseline itself is estimated from the data, it is
important to verify that the feature near
$\omega\simeq18.3$ is not produced by a particular choice
of baseline parametrization.

Our fiducial analysis uses ten equal-width bins in $x$ and a
cubic polynomial fit to the logarithm of the binned source
density, as described in Sec.~III.A.  We repeat the frequency
analysis while varying both the binning and the degree of the
low-order polynomial used to construct the smooth baseline.
The position of the dominant enhancement remains close to
$\omega\simeq18.3$ over these variations, while its power
shows only modest changes.

We therefore conclude that the location of the observed
feature is not strongly sensitive to the particular smooth
baseline parametrization adopted for $p_0(x)$.

\subsection{Five-fold cross-validation}

A more stringent test is obtained by separating the data used to
select the frequency from the data used to test it.  We therefore
perform a five-fold cross-validation analysis for the PRIMER--UDS
subsample.

The PRIMER--UDS sample is randomly divided into five approximately
equal subsets, or ``folds.''  In each realization, four folds,
containing about $80\%$ of the objects, are used as a training
sample.  The dominant frequency is determined using only this
training sample.  The remaining fold, containing about $20\%$ of
the objects, is then used as an independent test sample to evaluate
the statistical strength of the frequency selected from the
training data.

\begin{figure}[t]
\centering
\includegraphics[width=0.65\linewidth]{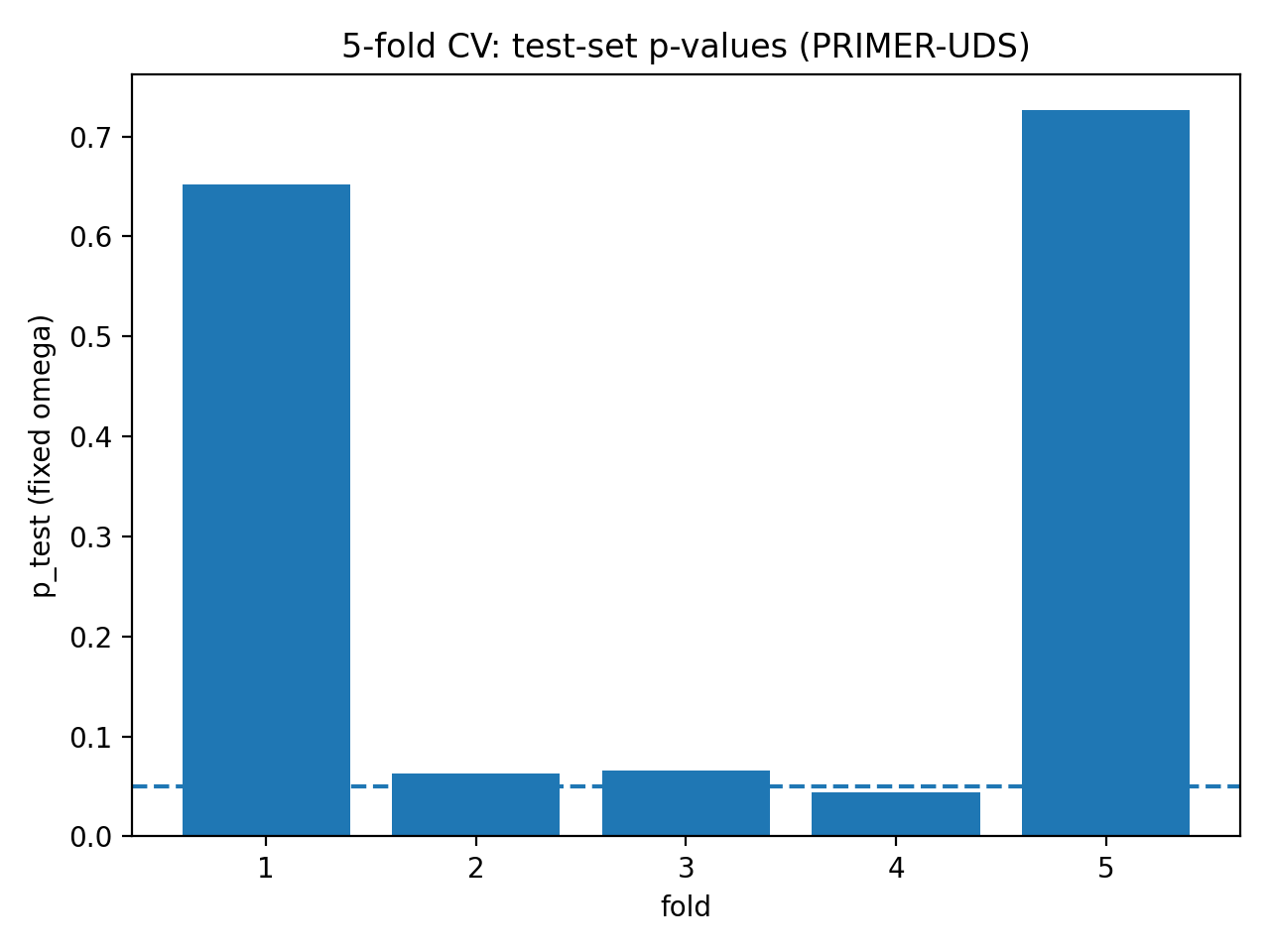}
\caption{
Test-set $p$-values obtained in the five-fold cross-validation
analysis of the PRIMER--UDS sample. In each fold, the dominant
frequency is selected using the training subset and then evaluated
on the held-out test subset. The dashed horizontal line indicates
the reference level $p=0.05$. The substantial variation among the
folds illustrates the limited statistical power of the reduced test
samples, while some folds retain relatively small test-set
$p$-values.
}

\label{fig:UDS_5fold_pvalues}
\end{figure}

 This procedure is repeated five times so that each fold serves once
as the independent test sample. Figure~\ref{fig:UDS_5fold_pvalues}
shows the resulting test-set $p$-values.  The horizontal axis labels
the five held-out folds, while the vertical axis gives the $p$-value
obtained in each independent test subset at the frequency selected
from the corresponding training sample.  The dashed horizontal line
marks $p=0.05$ as a reference level.  Thus, bars below this line
indicate folds in which the frequency selected from the training
sample retains relatively strong evidence in data that were not used
to determine that frequency.
The test-set significance varies substantially among the five folds,
as expected for the reduced sample size.  Some folds yield relatively
small $p$-values, whereas others do not show significant evidence
for the modulation.  The cross-validation test therefore provides
partial support for the stability of the feature, but also shows
that its statistical strength is sensitive to the particular
subsample used.

\subsection{Repeated holdout analysis}

The five-fold cross-validation test still depends on one particular
partition of the data.  To examine this dependence more directly,
we perform repeated random train/test splits of the PRIMER--UDS
sample.

In each realization, the dominant frequency is determined using
only the training subset.  This frequency is then held fixed and
its statistical significance is evaluated on the corresponding
independent test subset.  Repeating this procedure for many random
partitions produces a distribution of test-set $p$-values.

\begin{figure}[t]
\centering
\includegraphics[width=0.65\linewidth]{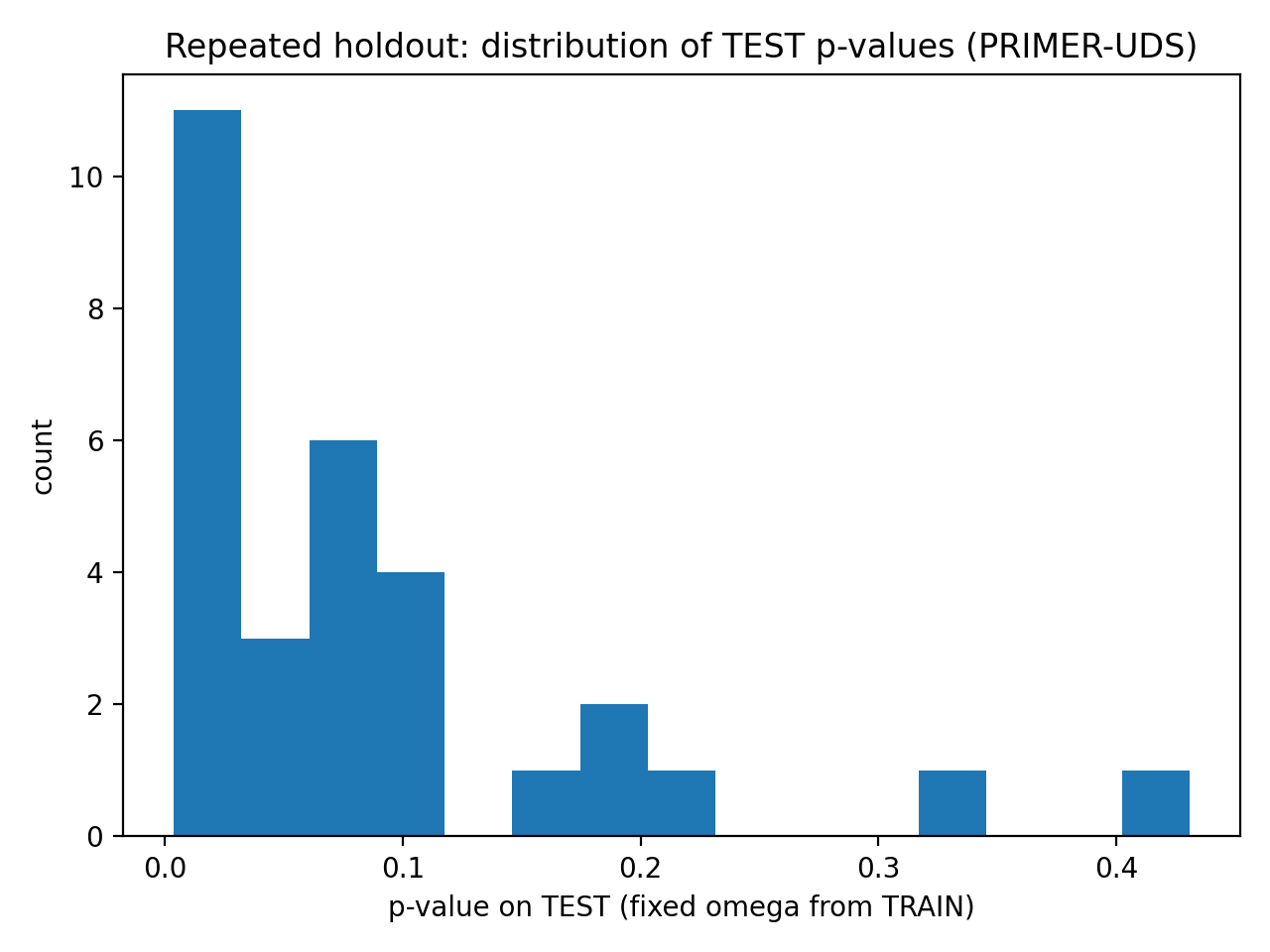}
\caption{
Distribution of test-set $p$-values obtained from repeated random
train/test splits of the PRIMER--UDS sample.  In each realization,
the dominant frequency is determined from the training subset and
then evaluated on the corresponding independent test subset.
The recurrence of low-$p$ realizations indicates that the result is
not tied to a particular partition of the data.
}
\label{fig:holdout}
\end{figure}

Figure~\ref{fig:holdout} shows this distribution.  The horizontal
axis represents the $p$-value obtained from the test subset, while
the vertical axis gives the number of holdout realizations in each
bin.  A concentration toward small $p$-values indicates that a
frequency selected from the training data tends to retain statistical
strength in data that were not used in selecting that frequency.

The distribution contains many realizations with relatively small
test-set $p$-values, although higher-$p$ cases are also present.
Thus the modulation is not equally strong for every possible
partition of the data.  Nevertheless, its recurrence over multiple
independent train/test splits indicates that the result is not tied
to one particular partition of the PRIMER--UDS sample.

\subsection{Field dependence}

The combined sample contains objects drawn from several independent
JWST survey fields, raising the possibility that the observed
enhancement could be dominated by a field-specific selection effect
or observational systematic.

We therefore examine the contribution of the individual survey
regions separately.  The PRIMER--UDS field provides a substantial
contribution to the total signal, in part because it contains a
relatively large number of objects.  However, similar enhancements
at compatible phase positions are also present in several
independent survey regions.

The modulation is therefore not confined to a single area of the
sky.  This disfavors, although it does not completely exclude, an
interpretation in which the feature is produced solely by a
field-specific observational bias or selection effect.

\subsection{Remaining observational systematics}

The tests above address the dependence of the result on the smooth
baseline, data partition, and survey field.  Several observational
systematics, however, cannot yet be fully quantified with the
present heterogeneous sample.

In particular, the catalog combines sources obtained with different
survey depths and selection criteria, and the adopted redshift
estimates are not of uniform quality.  Redshift uncertainties can
broaden or shift structure in $x=\ln(1+z)$, while differences between
photometric and spectroscopic redshift determinations may introduce
additional sample-dependent effects.  Possible contamination or
future reclassification of individual compact sources may also
modify the observed redshift distribution.

These effects do not by themselves demonstrate that the feature near
$\omega\simeq18.3$ is artificial, but neither can they be excluded
by the present analysis.  Recent studies have emphasized that the inferred redshift evolution
of the LRD population can depend sensitively on the adopted selection
criteria and on observational effects \cite{Rinaldi2026}. This issue is particularly relevant to the present analysis.  Our null distribution
$p_0(x)$ is introduced precisely to separate the smooth redshift
dependence of the observed population, including such selection-dependent
trends, from a possible oscillatory component in $x=\ln(1+z)$.
We stress, however, that this procedure does not by itself exclude an unidentified selection effect having structure on a scale comparable
to the observed modulation.  The peak near $\omega\simeq18.3$ should
therefore be regarded as evidence for a candidate log-periodic component
relative to the adopted null distribution, whose persistence must be
tested with larger and more homogeneous independent JWST samples.
\section{Possible physical interpretation}
\label{sec:interpretation}

If the observed modulation is confirmed with larger and more
homogeneous samples, the approximately constant spacing
\begin{equation}
 \Delta x=\Delta\ln(1+z)\simeq0.34
 \label{eq:preferred}
\end{equation}
would indicate a preferred multiplicative scale in $1+z$.

The relation between a recurrence interval in cosmic time and one
in logarithmic redshift follows directly from
\begin{equation}
 x=\ln(1+z)=-\ln a+\mathrm{const.},
\end{equation}
which gives
\begin{equation}
 \frac{dx}{dt}=-H(z).
\end{equation}
Thus, over a sufficiently short interval,
\begin{equation}
 |\Delta\ln(1+z)|
 \simeq H(z)\Delta t .
 \label{eq:deltat-deltax}
\end{equation}
A recurrence time that tracks a fixed fraction of the local
cosmological expansion time, $\Delta t\propto H^{-1}(z)$, would
therefore generate an approximately constant spacing in
$\ln(1+z)$.

A possible physical setting in which recurrent cosmological
dynamics can arise is a self-interacting Bose--Einstein condensate.
In the Fukuyama--Morikawa--Tatekawa framework, an attractive
self-interaction can drive repeated episodes of condensation,
collapse, and re-formation \cite{FukuyamaMorikawa2006,
FukuyamaMorikawaTatekawa2008,Fukuyama2026}.
The associated recurrence scale can be of cosmological relevance
and may lead to successive structure-formation episodes separated
approximately in logarithmic redshift.

The observed value,
\begin{equation}
 \Delta\ln(1+z)\simeq0.34,
\end{equation}
is of the same order as the logarithmic recurrence scale discussed
in this class of condensate models.  This numerical proximity is
suggestive, but it should not be interpreted as a parameter-free
prediction of the theory.  In particular, the recurrence scale
depends on the condensate dynamics and on its evolution in the
expanding background.

The statistical evidence for a possible log-periodic component and
its physical interpretation are therefore logically distinct.
Other mechanisms capable of generating an approximately preferred
scale in $\ln(1+z)$ should also be considered, and confirmation with
independent high-redshift samples is required before associating the
observed feature with a particular cosmological mechanism.

\section{Conclusions}
\label{sec:conclusions}

We have performed a model-independent search for coherent
log-periodic structure in the redshift distribution of a sample of
$N=341$ high-redshift compact sources.  Using the logarithmic
variable $x=\ln(1+z)$ and subtracting the Fourier contribution
expected from a smooth baseline distribution, we find a prominent enhancement at
$\omega_{\rm peak}=18.33$, corresponding to
$\Delta\ln(1+z)\simeq0.34$, or equivalently to a multiplicative
spacing of approximately $1.41$ in $1+z$.

At the location of the dominant enhancement,
$\omega_{\rm peak}=18.33$, the observed power is
$P_{\rm obs}=5.87$.  Monte Carlo catalogs generated from the
smooth null distribution give a local tail probability
$p_{\rm local}\simeq2.8\times10^{-3}$.  Accounting explicitly
for the frequency scan over the theoretically motivated
interval $16\leq\omega\leq21$, we obtain a global false-alarm
probability
\begin{equation}
 p_{\rm global}\simeq7.6\times10^{-3}.
\end{equation}
Thus the statistical significance is reduced by the
look--elsewhere correction, as expected, while the global
false-alarm probability remains below one percent within the
specified search interval.

Several robustness tests provide additional information on the
stability of the feature.  Its statistical strength is not strongly
sensitive to the adopted smooth baseline, while five-fold
cross-validation and repeated train/test holdout analyses show that
the signal survives in some independent subsets but with substantial
variation in significance.  Field-dependent tests further indicate
that the enhancement is not confined exclusively to a single survey
region.

The present evidence should therefore be regarded as suggestive
rather than conclusive.  Although the global false-alarm probability
remains below one percent within the theoretically motivated
frequency interval, confirmation will require larger and more
homogeneous high-redshift samples, together with improved control
of selection functions and redshift uncertainties.

If confirmed, an approximately constant spacing in
$\ln(1+z)$ would imply a preferred multiplicative scale in the
high-redshift source distribution.  Recurrent cosmological
condensate dynamics provide one possible physical interpretation,
but the statistical result itself does not depend on this mechanism.
Independent observational confirmation will be essential for
establishing whether the feature is physical and, if so, for
identifying its origin.

\end{document}